\documentclass[preprint,showkeys, preprintnumbers,amssymb]{revtex4}

\usepackage{graphicx}
\graphicspath{{figs/}}
\begin{document}
\title{A universal exponential factor in the dimensional crossover from graphene to graphite}
\author{Jin-Wu~Jiang}
\thanks{Electronic mail: phyjj@nus.edu.sg}
    \affiliation{Department of Physics and Centre for Computational Science and Engineering,
             National University of Singapore, Singapore 117542, Republic of Singapore }
\author{Jian-Sheng~Wang}
    \affiliation{Department of Physics and Centre for Computational Science and Engineering,
                 National University of Singapore, Singapore 117542, Republic of Singapore }

\date{\today}
\begin{abstract}
A universal exponential factor, $\gamma_{c}=\pi/2$, is disclosed for the dimensional crossover of few-layer graphene (FLG) from two-dimensional graphene to three-dimensional graphite. $\gamma_{c}$ is found by analyzing available experimental data on different properties of FLG with varying thickness. A theoretical study on the phonon spectrum of the vertical acoustic mode in FLG is carried out to further check this exponential factor $\gamma_{c}$. Interestingly, the same exponential factor appears in the dimensional crossover of the phonon mode. It turns out that the exponential factor $\gamma_{c}$ is related to the homogeneous Helmholtz-like molal equation in the mass transfer with a first order chemical reaction. The finding should provide valuable information for experimentalists and theorists in the future investigation on thickness dependent properties of FLG.
\end{abstract}

\maketitle

\pagebreak

\section{introduction}
The advancement of producing few-layer graphene (FLG) of high quality enables FLG to be a very promising candidate for future nano-devices. The single layer graphene sheet ($N=1$) is an one-atom-thick two-dimensional (2D) system; while FLG with $N\longrightarrow \infty$ is a three-dimensional (3D) graphite system. Hence, the FLG provides a ready-made platform for the investigation of dimensional crossover from 2D to 3D by increasing the layer number $N$. It is quite interesting to investigate the evolution for different physical properties of FLG during the dimensional crossover from 2D graphene to 3D graphite. In a very recent experiment,\cite{GhoshS} the in-plane thermal conductivity was measured for FLG with different $N$, and it was found that the dimensional crossover from 2D to 3D is very fast. For FLG with $N>4$, the value of the thermal conductivity is already very close to that of the 3D graphite. There are some other available experiments on the dimensional crossover of different aspects for FLG. Those experiments include the interface thermal conductance,\cite{Koh} the Raman G mode,\cite{Gupta} the second order D peak,\cite{Ferrari} the surface roughness,\cite{Lauffer} the electronic band gap,\cite{ZhouSY2007} the surface potential,\cite{Datta} etc. The main purpose of this work is to extract the common rules in the dimensional crossover of FLG from existing experiments, check them theoretically, and understand the underlying mechanism.

In this work, by analyzing various experimental data, we find a universal exponential factor $\gamma_{c}=\pi/2$ for the dimensional crossover of FLG from 2D graphene to 3D graphite. We then perform theoretical check on $\gamma_{c}$ by studying the phonon spectrum of the vertical acoustic (ZA) mode in FLG. We find that the phonon spectrum of the ZA mode varies exponentially during the dimensional crossover from 2D to 3D with exactly the same exponential factor $\gamma_{c}$. Finally, we give a physical interpretation for the value of $\gamma_{c}$ by relating it to the mass transfer phenomenon with a first order chemical reaction.

\section{analysis for experimental results}
In this section, we analyze several experiments on the dimensional crossover from 2D graphene to 3D graphite for different physical properties of FLG. These experiments cover thermal properties, electronic properties, Raman studies, and mechanical property of the FLG. We will demonstrate that all experimental data can be well described by exponential functions with a common exponential factor $\gamma_{c}=\pi/2$.

\begin{figure}[htpb]
  \begin{center}
    \scalebox{1.0}[1.0]{\includegraphics[width=8cm]{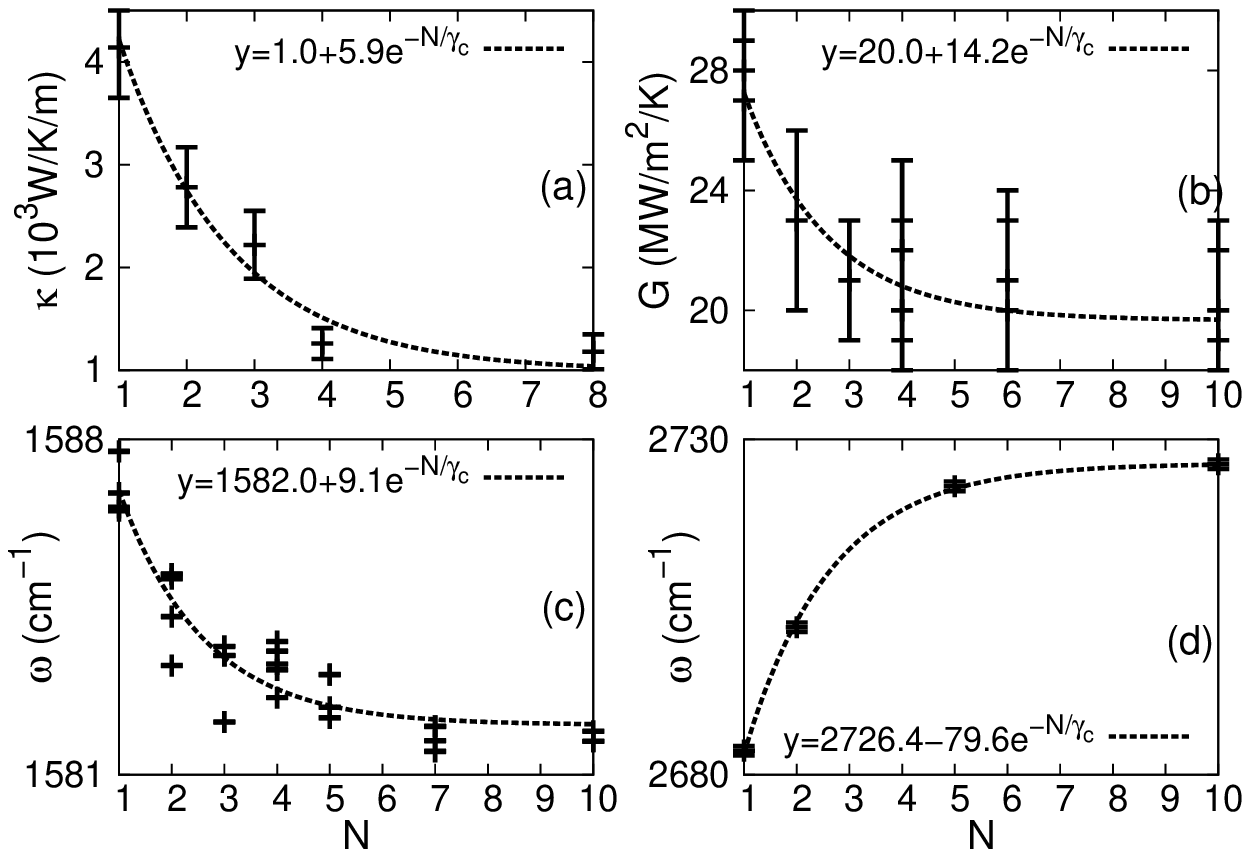}}
  \end{center}
  \caption{Experimental results on various physical quantities in FLG with different $N$. The constant $\gamma_{c}=\pi/2$ is the common exponential factor. (a) In-plane thermal conductivity $\kappa$ measured in Ref.~\onlinecite{GhoshS}. (b) Interface thermal conductance $G$ from Ref.~\onlinecite{Koh}. (c) The Raman G mode from Ref.~\onlinecite{Gupta}. (d) The second order D peak in Ref.~\onlinecite{Ferrari}.}
  \label{fig_exp1}
\end{figure}

Fig.~\ref{fig_exp1}~(a) shows the in-plane thermal conductivity of the FLG with different layer number $N$. Both the value and error bar in the figure are from Fig.2~(b) in Ref.~\onlinecite{GhoshS}. The temperature dependence of the G mode was used to detect the temperature gradient in the FLG. The obtained temperature gradient together with the thermal power of the laser can be used to derive the value of the thermal conductivity. This method has been applied in single layer graphene sheet by the same group and a thermal conductivity of 4000 W/m/K was discovered in this 2D system.\cite{Balandin} The thermal conductivity of the regular graphite is about 950 W/m/K. Fig.~\ref{fig_exp1}~(a) exhibits the dimensional crossover of the thermal conductivity from the 2D graphene to the 3D graphite systems. The thick FLG sample with $N=8$ has a larger uncertainty in its layer number with $6\le N \le 10$. The measured thermal conductivity is lower in thicker FLG samples. This result was attributed to the enhancement of the three phonon scattering channels by the inter-layer van de Waals (VDW) interaction. The experimental data can be fitted to an exponential function $\kappa=992.2+5916.9e^{-N/\gamma_{c}}$ with the exponential factor $\gamma_{c}=\pi/2$. In the limit of large $N$, $\kappa=992.2$ W/m/K yields the thermal conductivity of the 3D graphite within five percentage. The small exponential factor $\gamma_{c}=\pi/2$ indicates a fast crossover from 2D graphene to 3D graphite with the increase of $N$. It means that the thermal transport capability of FLG with $N=2$ is already very close to 3D graphite. In following, we will show that the same exponential factor $\gamma_{c}=\pi/2$ exists in other experiments.

Fig.~\ref{fig_exp1}~(b) is the vertical thermal conductance per unit area ($G$) v.s layer number $N$. These data are for 3.8 $\mu$m FLG samples from Fig.3~(b) in Ref.~\onlinecite{Koh}. The thermal conductance was measured by time-domain thermoreflectance method, which may give a 20$\%$ uncertainty in measured value. For each $N$, there are two values of $G$ corresponding to different experimental conditions. The low thermal conductance in the experiment was attributed to the Kapitza thermal resistance of various interfaces in this sandwich structure. Fig.~\ref{fig_exp1}~(b) shows that the thermal conductance decreases exponentially and reaches a saturate value quickly with the increase of layer number $N$. The data can be fitted to an exponential function, yielding the common exponential factor $\gamma_{c}$. This small exponential factor indicates that a FLG sample with $N=2$ can be used to mimic the 3D graphite quite reasonably.

Fig.~\ref{fig_exp1}~(c) shows the Raman study of the G mode in FLG with different $N$, which is from Fig.3 and Fig.4 of Ref.~\onlinecite{Gupta}. The error bar (about 0.01 cm$^{-1}$) is an estimation of possible errors introduced when data are read out from the experimental work. For each $N$, the variation in experimental frequencies is about 0.5 cm$^{-1}$, which was attributed to the spot-to-spot variation within the FLG films.\cite{Gupta} Fig.~\ref{fig_exp1}~(c) shows a red-shift of the G mode with increasing $N$. This red-shift is originating from the polarization of the bond charge due to the inter-layer VDW interaction of the FLG.\cite{JiangJW2008} The experimental data in the figure are fitted to an exponential function. We find that the common exponential factor $\gamma_{c}$ also presents itself in this experiment. In large $N$ limit, the exponential function gives 1582.04 cm$^{-1}$ for G mode. This value meets the frequency of G mode in 3D graphite (1581.91 cm$^{-1}$). This agreement confirms that the exponential function with common factor $\gamma_{c}$ is a good description of the experimental frequency of G mode.

The position of the second-order Raman D peak in FLG is shown in Fig.~\ref{fig_exp1}~(d). The data are read out from Fig.2 of Ref.~\onlinecite{Ferrari}. Similar experimental results can also be found in Ref.~\onlinecite{Calizo}. This frequency corresponds to the second-order Raman scattering of zone-boundary phonon around 2700 cm$^{-1}$. For the second-order Raman D peak, there is only a single component in graphene, while two components are found in graphite.\cite{Ferrari} We have extracted the position of this Raman peak by reading the value of the frequency at the center of the Raman peak. In the experiment, two Renishaw spectrometer were applied: 514 and 633 nm. We focus on the experimental data under 514 nm. Compared with results at 514 nm, the shape and the position of the Raman spectra at 633 nm show an overall shift, but the $N$-dependence is similar. As the Raman spectrum experiment is of high accuracy, the error bar (about 0.7 cm$^{-1}$) for each data in Fig.~\ref{fig_exp1}~(d) is estimated from possible errors during we reading out data from Ref.~\onlinecite{Calizo}. This figure shows that the frequency of the 2D mode is well described by $\omega=2726.4-79.6e^{-N/\gamma_{c}}$, where the constant 2726.4 cm$^{-1}$ is exactly the value in 3D graphite. Interestingly enough, the dimensional crossover of the frequency is determined by the common exponential factor $\gamma_{c}=\pi/2$, the same as previous experiments.

\begin{figure}[htpb]
  \begin{center}
    \scalebox{1.0}[1.0]{\includegraphics[width=10cm]{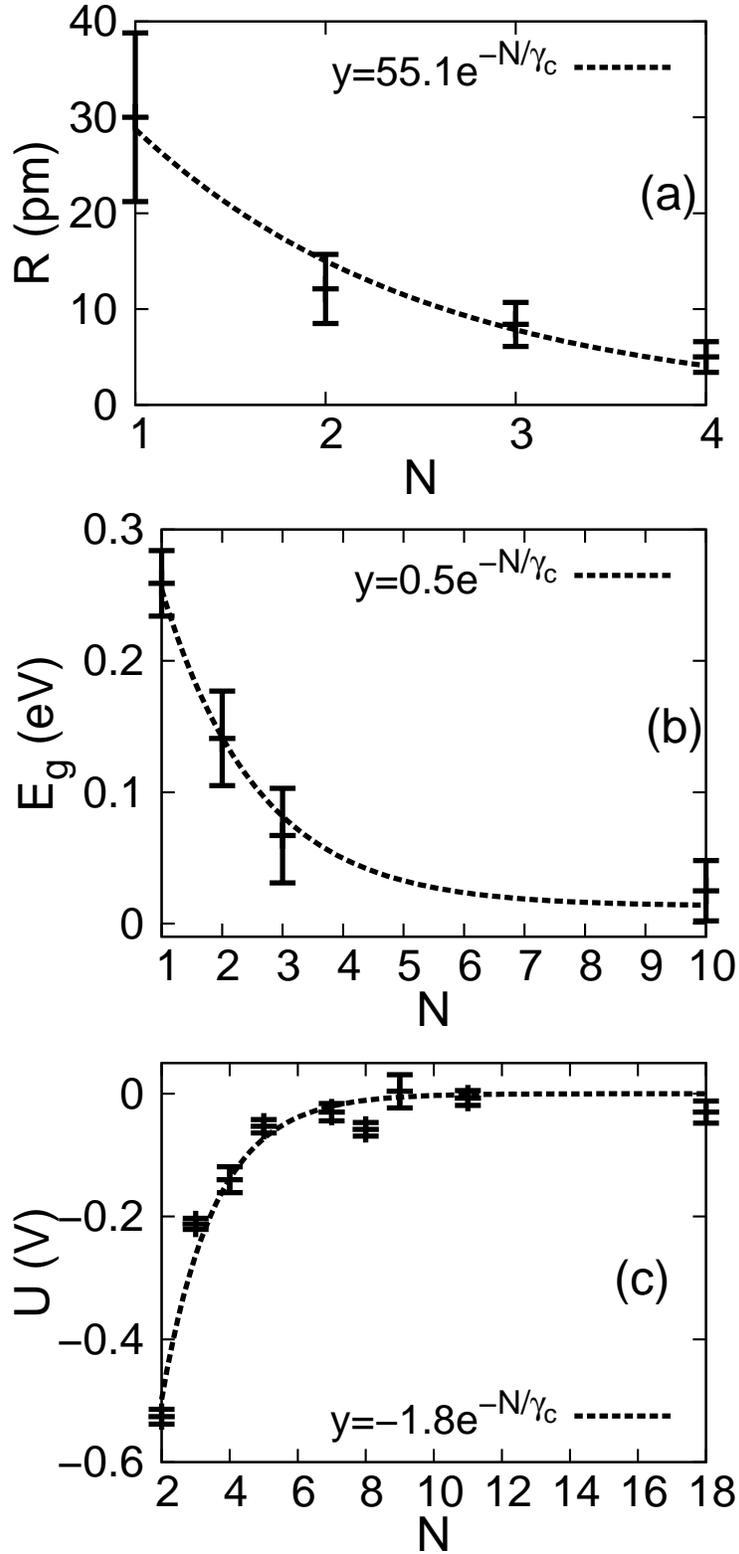}}
  \end{center}
  \caption{Experimental results for the dimensional crossover from 2D to 3D of three physical quantities in FLG. The constant $\gamma_{c}=\pi/2$ is the common exponential factor. (a). The surface roughness of FLG.\cite{Lauffer} (b). The electronic band gap.\cite{ZhouSY2007} (c). The surface potential of FLG.\cite{Datta}}
  \label{fig_exp2}
\end{figure}

As shown in Fig.~\ref{fig_exp2}~(a) is the roughness of FLG with different thickness. This experiment was performed by Lauffer {\it et~al.} in Ref.~\onlinecite{Lauffer}. The interface-induced roughness $R$ reflects the height variations of the FLG. The error bars in the figure are mainly caused by the variations in the tip geometry. The single layer graphene sheet has a rather large roughness $R$=30pm, and $R$ decreases quickly with the increase of $N$. Similar roughness of single layer graphene sheet was also observed by high-resolution atomic force microscopy.\cite{LiuCH} The decrease of the roughness with increasing thickness was also found by Brar {\it et~al.}\cite{Brar}, where the Bernal stacking structure of FLG was observed. This experiment demonstrates that thicker FLG samples are more smooth due to the inter-layer VDW interactions. The roughness $R$ can be fitted to exponential function $R=55.1e^{-N/\gamma_{c}}$. This function yields a zero roughness for 3D graphite. Again, the common exponential factor $\gamma_{c}$ exists in this experiment.

Fig.~\ref{fig_exp2}~(b) shows the electronic band gap v.s $N$ in FLG. The angle-resolved photo emission spectroscopy was taken for FLG to deduce the electronic band gap in Ref.~\onlinecite{ZhouSY2007}. A band gap of 0.26 eV is obtained for single layer graphene sheet, which benefits its application in electronic devices field. The value of $E_{g}=0.025$eV for $N=10$ is actually the band gap in 3D graphite which was estimated from a band-structure calculation.\cite{McClure,DresselhausMS} This value should be set as the band gap for infinite large $N$. However, we find that $N=10$ is already large enough. There is no difference in the fitting function if we set this value as the band gap for FLG with $N=100$. As shown in the figure, the band gap decreases exponentially with the increase of layer number $N$. The common factor $\gamma_{c}$ appears in this experiment as the exponential factor. Although the existence of this band gap is widely accepted, the origin of the gap is a current object of an ongoing debate. Zhou {\it et~al.} interpreted this band gap as resulting from the breaking of sublattice symmetry due to the graphene substrate interaction in the epitaxially grown FLG samples on SiC substrate;\cite{ZhouSY2008} while Rotenberg {\it et~al.} believed that the origin of the band gap is due to the modulation of the lateral structure of the FLG samples.\cite{RotenbergE} A more definitive interpretation of the opening has been presented by Bostwick {\it et~al.}, where a renormalization of the bands around the Dirac crossing is caused by the electron-plasmon coupling.\cite{BostwickA}

The surface potential of FLG with different $N$ is displayed in Fig.~\ref{fig_exp2}~(c). These experimental data are reported in Fig.3 of Ref.~\onlinecite{Datta}. Similar results can also be found from Ref.~\onlinecite{Ohta}. The surface potential is the electrostatic potential measured at the exterior layer of the FLG sample. The value was probed by the electrostatic force microscopy. It was proposed that the surface potential is affected by the charge exchange between the FLG and the substrate. We can see that the evolution of the surface potential with $N$ can be well described by the exponential function, and we obtain the common exponential factor $\gamma_{c}$.

\section{dimensional crossover for ZA phonon mode}
\begin{figure}[htpb]
  \begin{center}
    \scalebox{1.25}[1.25]{\includegraphics[width=7cm]{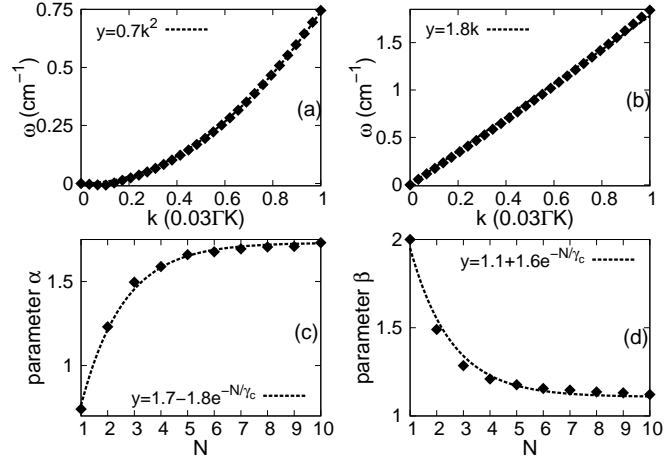}}
  \end{center}
  \caption{Phonon dispersion of ZA mode in FLG is $\omega=\alpha k^{\beta}$. (a). ZA mode in graphene sheet is a flexure mode with parabolic spectrum, $\beta=2.0$. (b). ZA mode in 3D graphite is linear, $\beta=1.0$. (c). The coefficient $\alpha$ v.s $N$. (d). The power factor $\beta$ v.s $N$.}
  \label{fig_fm}
\end{figure}

In the above, a universal exponential factor $\gamma_{c}=\pi/2$ is revealed by analyzing various experimental results on the dimensional crossover from 2D graphene to 3D graphite. The rest of this paper is devoted to the discussion of the dimensional crossover for the ZA mode. ZA mode is the acoustic phonon in the vertical direction of the graphene. This is the so-called flexure mode in graphene which has a quadratic phonon spectrum due to the 2D plane sheet structure of graphene.\cite{Born,Landau} We study the evolution of the ZA mode during the dimensional crossover from 2D graphene to 3D graphite. The phonon dispersion of ZA mode can be described by a formula $\omega=\alpha k^{\beta}$, where $\alpha$ and $\beta$ are two parameters. The phonon dispersion is obtained by solving the lattice dynamics of FLG at the equilibrium structure from the ``General Utility Lattice Program" (GULP)\cite{Gale}. The in-plane carbon-carbon interaction is described by the Brenner potential.\cite{Brenner} For the inter-layer interaction, since the distance between two layers is out of the interaction range of Brenner potential, we introduce Lennard-Jones potential, $V(r)=4\epsilon ((\sigma/r)^{12}-(\sigma/r)^{6})$, with $\epsilon=2.5$ meV and $\sigma=3.37$~{\AA}. The value of the length parameter $\sigma$ is fitted to the space between adjacent layers in three-dimensional graphite\cite{SaitoR} as 3.35~{\AA}. The energy parameter $\epsilon$ is fitted to the phonon dispersion along $\Gamma A$ direction in graphite.\cite{Nicklow} The cutoff for Lennard-Jones potential is chosen as 20~{\AA}, which is large enough; i.e., all results do not vary with further increasing this cutoff value. Fig.~\ref{fig_fm}~(a) is the ZA mode for $N=1$, i.e single layer graphene sheet. The ZA mode in graphene is a perfect flexure mode with $\beta=2$. We have considered the long wave phonon with the wave vector from $\Gamma$ to 0.03$\Gamma K$ in the Broullin zone, where $\Gamma K=4\pi/(3\sqrt{3}b)$ and the bond length $b=1.42$~{\AA}. Panel (b) is the ZA mode in 3D graphite, where $\alpha=1.8$ and $\beta=1.0$. Panels (c) and (d) display the two parameters $\alpha$ and $\beta$ as a function of layer number $1\le N \le 10$. These two curves can be fitted to exponential functions, and the exponential factor $\gamma_{c}=\pi/2$ is the same as what we have discovered from experiment analysis. In the limit of large $N$, $\alpha=1.72$ and $\beta=1.07$ give the results for 3D graphite within some small errors.

We point out that the difference in ZA mode for different $N$ originates from the two nonequivalent sublattices in the honeycomb structure of graphene. In the elastic model, the FLG are constructed by many uniform elastic plates coupled through VDW interactions.\cite{Kitipornchai} The solution for the ZA mode is $u_{n}(x,y)=Ae^{i(k_{x}x+k_{y}y)}e^{i\omega t}$, which is independent of the layer index $n$. This solution simply indicates that all graphene layers vibrate with the same amplitude and phase. As a result, the VDW has no contribution to the ZA mode, and the phonon dispersion of the ZA mode in FLG should be the same as single layer graphene, i.e $\omega=0.7*k^{2}$ cm$^{-1}$. However, the elastic model loses all lattice structure information, especially the nonequivalent property of the two lattice sites $A$ and $B$. In the Bernal-stacked FLG, two adjacent layers have a relative shift by a carbon-carbon bond length. This relative shift can be considered by modifying the solution of the ZA mode to be $n$-dependent: $u_{2n}(x,y)=Ae^{i(k_{x}x+k_{y}y)}e^{i\omega t}$ for even layer and $u_{2n+1}(x,y)=Ae^{i(k_{x}x+k_{y}y)}e^{i\omega t}e^{i\vec{k}\cdot \vec{b}}=u_{2n}e^{i\vec{k}\cdot \vec{b}}$ for odd layer. The vector $\vec{b}$ is the relative shift between two adjacent layers. This solution exhibits that the two adjacent layers have a phase difference during vibration, which has effect on the long wave length phonon. As a result, the VDW interaction can affect the phonon dispersion of the ZA mode; thus the ZA mode is different in FLG with different $N$.

\section{underlying mechanism}
The above universal exponential factor $\gamma_{c}=\pi/2$ is probably rooting in the stacking configuration structure of the FLG, as this constant exists in quite different physical quantities. Let us consider a general quantity $Q$ siting on one carbon atom in the first layer of the FLG. There is a possibility for $Q$ to hop to other sites. We focus on the hoping phenomenon in the vertical ($z$) direction of FLG. Under the first-nearest-neighboring tight-binding approximation, the hopping parameter $K_{0}$ in $z$ direction is inverse proportional to the inter-layer space $c$.\cite{MahanGD} In the meantime, larger in-plane bond length $b$ will result in larger $K_{0}$, because in-plane hoping is suppressed in case of larger $b$ which enhances hoping in $z$ direction. These two effects determine the hoping parameter as $K_{0}=b/c$. As a result, the hopping rate in $z$ direction is $g_{Q}'=-K_{0}Q$ under the lowest-order approximation. Before writing down the equation for $Q$, it is important to point out that the hopping process discussed here is actually analogous to the mass transfer phenomenon with a first order chemical reaction.\cite{DasSK} The quantity $Q$ corresponds to the distribution of molecule concentration. Hoping in $z$ direction is equivalent to the chemical reaction to absorb the molecule. The hoping rate $g_{Q}'=-K_{0}Q$ is corresponding to the absorption rate of molecule under the first-order chemical reaction with $K_{0}$ as the reaction constant. By analogy with mass transfer, the distribution of quantity $Q$ along $z$ axis in steady state is governed by the molal equation:\cite{DasSK}
\begin{eqnarray}
\frac{d^{2}Q}{dN^{2}}-K_{0}Q=0,
\end{eqnarray}
where the coordinate $z$ is discretized in FLG system. The solution for this homogeneous Helmholtz-like equation under vanishing boundary condition is:\cite{Polyanin} $Q\propto e^{-N/\gamma_{c}}$ with $\gamma_{c}=\sqrt{1/K_{0}}$. This exponential solution is the origin of previous exponential functions in the dimensional crossover of different physical quantities; and the exponential factor $\gamma_{c}=\sqrt{1/K_{0}}=\sqrt{c/b}=1.54$ agrees quite well with above $\gamma_{c}=\pi/2=1.57$ from experimental and theoretical analysis. It should be noted that the we are studying the Bernal stacking FLG sample in our work. For turbostratic graphite or epitaxial graphene grown on SiC(000-1) with a peculiar turbostratic, non Bernal stacking, many of its properties are strongly affected. For example, the Raman G modes undergo stronger red-shift in AAAA stacking FLG than that of the Bernal stacking FLG;\cite{JiangJW2008} thus the exponential factor is slightly smaller than $\pi/2$. In our model, we have obtained the hoping parameter within the first order approximation. This approximation is more reasonable for Bernal stacking structure, where only one of the two sublattice has inter-layer neighbors.

\section{conclusions}
To summarize, we find a universal exponential factor $\gamma_{c}=\pi/2$ during the dimensional crossover of FLG from 2D graphene to 3D graphite. We obtain this exponential factor through the analysis of several experiments on the transition of different physical properties of the FLG with increasing layer number $N$. We then study theoretically the ZA mode in FLG with phonon spectrum $\omega=\alpha k^{\beta}$. We find that the two coefficients, $\alpha$ and $\beta$, vary exponentially with the increase of $N$, and the exponential factor is exactly the same as what we have obtained from the analysis of experiments. We also demonstrate that $\gamma_{c}$ is related to the mass transfer phenomenon with a first order chemical reaction.

\textbf{Acknowledgements} We thank D. G. Cahill for sharing their experimental data with us prior to publication (Ref.~\onlinecite{Koh}). The work is supported by a Faculty Research Grant of R-144-000-257-112 of National University of Singapore.

\end{document}